\documentclass[journal]{IEEEtran}

\usepackage[normalem]{ulem}
\usepackage{cancel}
\usepackage[dvipsnames]{xcolor}
\usepackage{hyperref}
\usepackage{epsfig,epsf,epstopdf}
\graphicspath{ {./Figs/} }
\usepackage[cmex10]{amsmath}
\usepackage{amsfonts}
\usepackage{amssymb}
\usepackage[noend]{algpseudocode}
\makeatletter
\def\BState{\State\hskip-\ALG@thistlm}
\makeatother
\usepackage{cite}
\usepackage{paralist}
\usepackage{color}
\usepackage{array}

\definecolor{blue}{rgb}{0, 0.1, 0.8}
\usepackage{mathrsfs}
\usepackage{multirow}
\usepackage{subfigure}
\usepackage{graphicx}
\usepackage{caption}
\usepackage[font=small,justification=justified,singlelinecheck=false]{caption}

\usepackage{nomencl}
\usepackage{framed}

\ifCLASSINFOpdf
 
\else
 
\fi

\makenomenclature
\begin{document}

\title{Secondary Defense Strategies of AC Microgrids Against Generally Unbounded Attacks}

\author{Yichao Wang, Mohamadamin Rajabinezhad and Shan Zuo,~\IEEEmembership{Member,~IEEE}% <-this % stops a space
\thanks{Yichao Wang, Mohamadamin Rajabinezhad
 and Shan Zuo are with the Department of Electrical and Computer Engineering, University of Connecticut, Storrs, CT, 06269 USA (e-mail: yichao.wang@uconn.edu; mohamadamin.rajabinezhad@uconn.edu; shan.zuo@uconn.edu). 
 
 This work has been submitted to the IEEE for possible publication. Copyright may be transferred without notice, after which this version may no longer be accessible.
}
}

\maketitle

\begin{abstract}
This paper develops a fully distributed attack-resilient secondary defense strategies for AC microgrids, addressing more generally unbounded attacks on control input channels than those addressed in existing literature. The secondary control of local inverter includes consensus-based voltage and current regulators utilizing relative information from neighboring inverters. This distributed control approach relies on localized control and a sparse communication network, making it susceptible to malicious cyber-physical attacks that can impair consensus performance and potentially destabilize the overall microgrid. 
In contrast to existing solutions that are limited to addressing either bounded faults, noises or unbounded attacks with bounded first-order time derivatives, we aim to surpass these constraints and enhance the defense capabilities of counteracting cyber-physical attacks by enabling the AC microgrids adopting the proposed strategies to withstand a much wider range of unbounded cyber-attack signals. Fully distributed attack-resilient secondary defense strategies are developed for AC microgrids to counteract the detrimental effects of generally unbounded attacks on control input channels. Rigorous proofs using Lyapunov techniques demonstrate that the proposed defense strategies accomplish the uniformly ultimately bounded convergence on frequency regulation and achieve voltage containment and active power sharing simultaneously for multi-inverter-based AC microgrids in the face of generally unbounded attacks. The proposed defense strategies are validated on a modified IEEE 34-bus test feeder benchmark system incorporating four inverter-based DERs. 
\end{abstract}

\begin{IEEEkeywords}
Containment, inverters, microgrids, resilience, unbounded attacks.           
\end{IEEEkeywords}

\section{Introduction}
Microgrids are emerging as a promising solution for the development of distributed energy resources (DERs). These systems must operate in a coordinated and efficient manner to ensure grid stability and reliability. The utilization of droop control is widespread in inverter-based DERs to ensure power sharing. However it often leads to undesirable frequency and voltage deviations. To mitigate this issue and coordinate multiple inverter-based DERs, distributed secondary control is seen as an effective strategy due to its flexibility, scalability, and robustness, where the communication network assumes a pivotal role in achieving the expected functionalities. Distributed cooperative secondary control of AC microgrids relies on consensus and containment approaches to accomplish frequency regulation \cite{yz052801} and voltage containment \cite{yz052802}. These approaches depend on local sensing and neighborhood relative information through sparse communication networks \cite{yz052804,wang2022distributed,liang2016review}.

% Microgrids are increasingly being recognized as a potent solution for the advancement of Distributed Energy Resources (DERs). It is imperative that these systems function in a coordinated and efficient manner to guarantee grid stability and reliability. Droop control is ubiquitously employed in inverter-based DERs to facilitate power sharing, albeit it frequently engenders undesirable deviations in frequency and voltage. To mitigate this issue and orchestrate multiple inverter-based DERs, distributed secondary control is perceived as an efficacious strategy, attributed to its flexibility, scalability, and robustness. In this context, the communication network assumes a pivotal role in realizing the anticipated functionalities. Distributed cooperative secondary control of AC microgrids leverages consensus and containment methodologies to achieve frequency regulation \cite{yz052801} and voltage containment \cite{yz052802}, respectively. These methodologies hinge on local sensing and neighborhood relative information through sparse communication networks \cite{yz052804,wang2022distributed,liang2016review}.

Microgrids are cyber-physical systems integrating the physical layer of power electronics and the distribution network and the cyber layer of distributed local controllers and sparse communication networks\cite{yz052805,yz052807}. They are under the potential threat of cyber-physical attacks that may occur in a coordinated fashion, targeting various components such as sensors, actuators, and communication channels to undermine the overall performance and stability of the microgrid \cite{tadepalli2022distributed}. For instance, attackers could disrupt the synchronization of inverter-based resources within microgrids by manipulating their output frequency and voltage through malicious false data injection (FDI) attacks. The existing research on microgrids has mainly focused on two approaches employed to counteract cyber-physical attacks. The first approach involves the detection of compromised agents, followed by recovery or isolation strategies \cite{mustafa2019detection,yz053101,tan2022false}. This approach often necessitates stringent criteria concerning the quantity of compromised agents and/or the graphical connectivity of the communication network. Furthermore, the central premise of the “detect and recover or isolate” method is that the attack signals are detectable. Whereas, stealthy attacks launched by intelligent attackers are generally undetectable. Since stealthy FDI attacks could bypass existing attack-detection algorithms for power systems \cite{yz052803,yc600,xiao2022three}.

Nonetheless, protection and prevention against stealthy and intelligent attackers are not always possible using attack-detection methods, and a paradigm shift to enhance the self-resilience of the large-scale networked microgrids by developing attack-resilient control protocols is the overarching objective for safeguarding the nation's critical infrastructures.
Distributed resilient control protocols are able to maintain an acceptable level of performance by mitigating the adverse impact of external perturbations and attacks that are propagated within the system. The primary focus is to devise locally distributed control approaches to enhance the self-resilience of the microgrids against malicious attacks without detecting and identifying the corrupted agents \cite{yz053102,yz05261815,yz052911,yz052912,shi2021observer,sadabadi2021fully,yang2023distributed,liu2021robust,sahoo2020resilient,zuo2022adaptive}.

In Ref \cite{shi2021observer}, a resilient control method based on observation is proposed. This approach enhances the conventional distributed control strategy by incorporating compensation terms incorporating errors between neighboring frequency and active power signals. Ref \cite{sadabadi2021fully} has proposed a resilient secondary frequency control for AC microgrids based on a resilient index to overcome FDI attacks. The authors in \cite{yang2023distributed} have proposed a technique that involves two stages. The first stage involves designing a distributed convergent observer, which is based on an iterative estimation technique for estimating the attack signals. This estimation is then used in the second stage to design a distributed resilient
controller in the secondary control layer. Ref \cite{liu2021robust} has introduced a robust and resilient distributed optimal
frequency control in AC microgrids, which is facilitated by integrating the original cyber-physical system with an auxiliary layer of the communication network. The resilient control strategy suggested in reference \cite{sahoo2020resilient} to lessen the impact of potential cyber-attacks is designed to maintain the functionality of the grid system even if $N-1$ (in a system with $N$ inverters) are subjected to FDI attacks.

\begin{figure}[!t]
\centering
\includegraphics[width=3.5in]{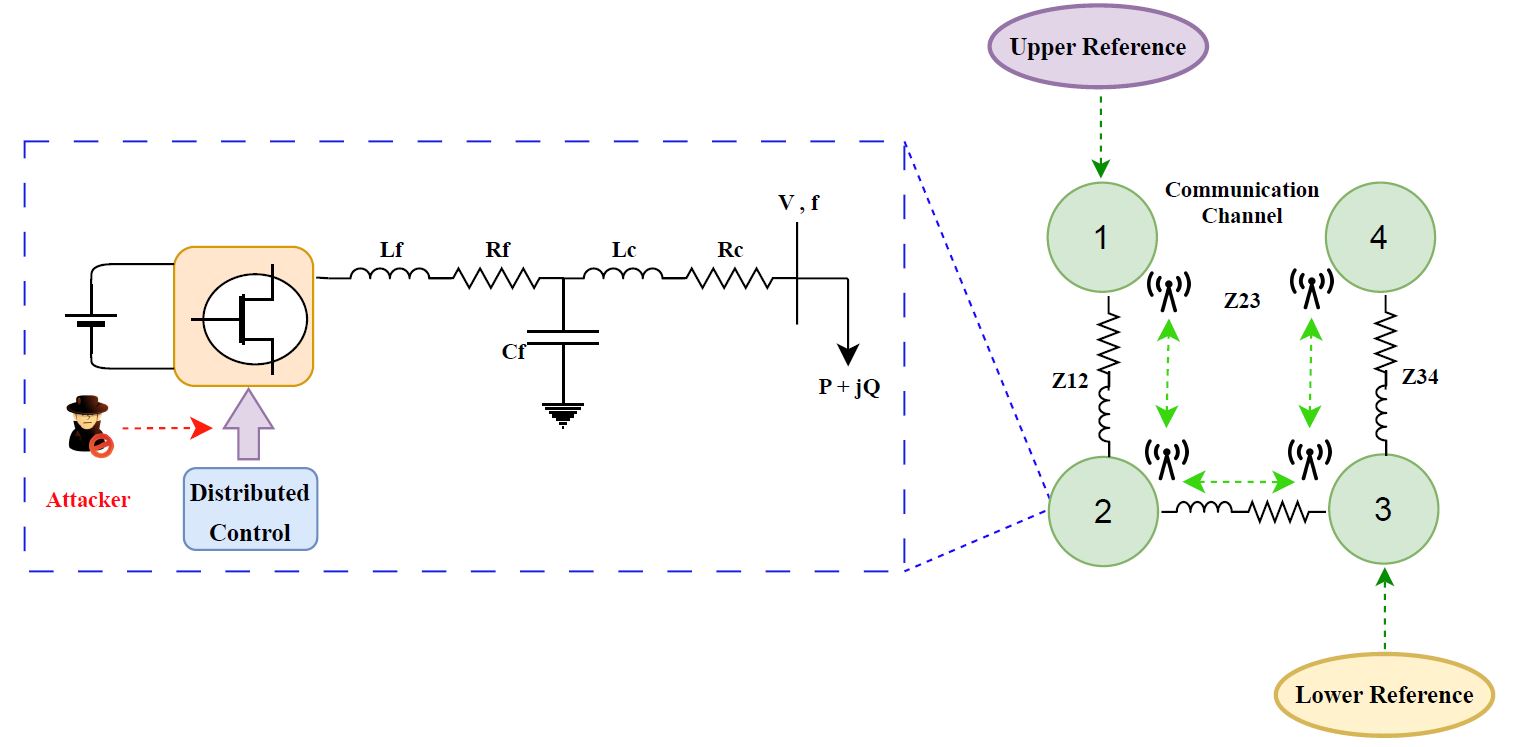}
\caption{A networked multi-inverter system under actuator attacks.}
\label{FIG1}
\end{figure}

It is worth noting that in the existing literature, disturbances, noises, faults, or attacks have generally been treated as bounded signals, with a few exceptions, such as those in the studies by\cite{liu2023resilient,yz052802,zuo2022adaptive,zhou2023distributed}. However, in reality, attackers may intentionally launch unbounded injections at the cyber-physical system to maximize their damage. Therefore, it is crucial to develop cyber-physical defense strategies that can withstand generally unbounded attacks to ensure the reliability and security of AC microgrids. In this paper, we address the practical yet challenging case of cooperative secondary control of AC microgrids in the face of generally unbounded attacks, as illustrated in Figure~\ref{FIG1}. Specifically, we consider a larger group of unbounded cyber-physical attacks injected into the control input channels that can be mitigated, with a relaxed constraint that only the high-order time derivative ($\gamma^{th}$) of the attack signal is required to be bounded. The contributions of this paper are summarized as follows:

$\bullet$ Fully distributed attack-resilient defense strategies are proposed for the secondary frequency and voltage control of AC microgrids. Specifically, a compensational signal is designed to counteract the detrimental effect of generally unbounded cyber-physical attacks on control input channels by incorporating an adaptively tuned parameter that reflects the neighborhood relative information. In contrast to existing solutions in \cite{liu2023resilient,yz052802,zuo2022adaptive,zhou2023distributed}, which are only capable of addressing a limited range of unbounded attack signals whose first-order time derivatives are required to be bounded, the proposed cyber-physical defense strategies handle more generally unbounded attack signals with a more relaxed constraint of bounded high-order time derivatives. This relaxation enables the AC Microgrids adopting the proposed strategies to withstand a much wider range of unbounded cyber-physical attack signals and, hence, enhance the the defense capability of AC microgrids against malicious cyber-physical attacks.
%{\color{red}This relaxation of the assumption that the first-time derivative of the cyber-attack signals' magnitude is bounded makes the control framework applicable to a wider range of cyber-attack signals and capable of dealing with devastating cyber-attacks.} as a result of relaxing the constraint on attack signals the defense strategies are capable of dealing with

%whose magnitude is up to ${\kappa_i} {t^\gamma }$, where $\kappa_i$ and $\gamma$ are positive numbers, and $t$ is time. 

$\bullet$ Rigorous proof based on Lyapunov stability analysis is provided to show that the proposed cyber-physical defense strategies achieve uniformly ultimately bounded (UUB) convergence for frequency regulation, voltage containment and power sharing against generally unbounded attacks on control input channels. This confirms that the proposed provably correct defense strategies successfully mitigate the aggregated adverse effects caused by generally unbounded attacks. Moreover, the ultimate bounds of frequency and voltage of each inverter can be tuned by properly adjusting the adaptation gains in the adaptive tuning laws. 

$\bullet$ The proposed cyber-physical defense strategies are fully distributed without requiring any global information and, hence, are scalable with plug-and-play capability. The efficacy and resilience of the proposed attack-resilient secondary defense strategies are validated through comprehensive case studies on a modified IEEE 34-bus test feeder benchmark system.

The rest of the paper is structured as follows: Section II reviews the preliminaries on graph theory and introduces the notations used in this paper. Section III presents the conventional cooperative secondary control of AC microgrids. In Section IV, the attack-resilient defense problems for secondary frequency and voltage control are formulated under generally unbounded attacks on control input channels. Fully distributed attack-resilient defense strategies are developed in section V. Section VI validates the enhanced resilience of the proposed attack-resilient defense strategies with comparative case studies on a modified IEEE 34-bus test feeder benchmark system incorporating four inverter-based DERs. Finally, Section VII concludes the paper.

%Alternatively, this letter explores adaptive techniques to address unknown unbounded attacks on input signals of the control loops, which are referred to as the actuator attacks. This letter considers the unbounded actuator attacks on both frequency and voltage control loops of an inverter, as illustrated in Fig.~\ref{FIG1}, which could severely destabilize the synchronization mechanism among microgrid inverters. The contributions of this letter are two-fold:

%Our architecture is more general which can be used both in multi-AC microgrids and DC microgrids such power systems.

%$\bullet$ A resilient control method is proposed for both secondary frequency and voltage control loops in the face of unknown unbounded actuator attacks. Compared to the observer-based techniques in \cite{yz052802}, this control method does not need addtional cyber layers for information exchange among observers, offering reduced computational complexity and system vulnerability to cyber attacks.

%$\bullet$ Stability analysis using Lyapunov techniques show that the proposed method is resilient to unbounded actuator attacks by preserving the uniformly ultimately bounded (UUB) consensus for frequency regulation and voltage containment, respectively. Moreover, the ultimate bound can be set by adjusting the tuning parameters. That is, the frequency and voltage terms can be tuned to converge to an arbitrarily small neighborhood around their respective reference values.

%The rest of this letter is organized as follows: Preliminaries on graph theory and notations are given in Section II. Section III reviews the conventional cooperative secondary control of AC microgrids. Section IV formulates the attack-resilient frequency and voltage control problems. Distributed resilient controller design is discussed in Section V. The efficacy of the proposed control method is verified for an AC microgrid in Section VI. Section VII concludes the letter.

\section{Preliminaries on Graph Theory and Notations}

Consider a communication network with $N$ inverters and two leader nodes, represented by a time-invariant weighted digraph $\mathscr{G}$. This digraph is associated with an adjacency matrix $\mathcal{A}= [{a_{ij}}] \in {\mathbb{R}^{N \times N}}$. Define ${\mathcal D}= \operatorname{diag}({d_i}) \in {\mathbb{R}^{N \times N}}$ and ${\mathcal L}= {\mathcal D} - {\mathcal A}$ as the in-degree matrix and the corresponding Laplacian matrix, respectively, where ${d_i} = {\sum\nolimits_{j = 1}^N}{a_{ij}}$. The two leader nodes issue the upper and lower reference values, respectively. $g_{ik}$ is the pinning gain from the $k^{th}$ leader to the $i^{th}$ inverter, brought together in the diagonal matrix ${\mathcal{G}_k} = \operatorname{diag} \left( {g_{ik}} \right)$.

${\sigma _{\min }}( \cdot )$, and ${\sigma _{\max }}(\cdot)$ are the minimum and maximum singular values of a given matrix, respectively. $\mathscr{F}$ and $\mathscr{L}$ denote the sets of $\left\{ {1,2,...,N} \right\}$ and $\left\{ {N + 1,N + 2} \right\}$, respectively. ${{\mathbf{1}}_N} \in {\mathbb{R}^N}$ is a column vector where all entries are one. $ \otimes $, $\operatorname{diag} \{  \cdot  \}$, $\left\|  \cdot  \right\|$, and $\left|  \cdot  \right|$ denote the Kronecker product, a block diagonal matrix, Euclidean norm of a given vector, and the absolute value of a given scalar, respectively. $f(x)\in C^n \Leftrightarrow f^{(n)}(x)\in C$, where $C$ is the class of continuous functions\cite{widder1989advanced}.

\section{Conventional Cooperative Secondary Control of AC Microgrids}
Conventional secondary control acts as an actuator by providing the input control signals for tuning the setpoints of decentralized primary control. The primary droop mechanism is given as follows for the $i^{th}$ inverter
\begin{align}
& \omega_i = \omega_{n_i}-m_{P_i}P_i,
\label{eq1}\\
& v_{odi} = {V_{n_i}} - {n_{Q_i}}{Q_i},
\label{eq2}
\end{align}
where $P_i$ and $Q_i$ are the active and reactive powers, respectively. $\omega _i$ and $v_{odi}$ are the operating angular frequency and the $d$ component of in $abc$ to $dq0$ transform (park transform) of inverter terminal voltage, respectively. ${\omega _{n_i}}$ and ${V_{n_i}}$ are the setpoints for the primary droop mechanisms fed from the secondary control layer. ${m_{P_i}}$ and ${n_{Q_i}}$ are $P-\omega$  and $Q-v$ droop coefficients selected per inverters' power ratings. 

We differentiate the droop relations in \eqref{eq1} and \eqref{eq2} with respect to time to obtain
\begin{align}
& {{\dot \omega }_{n_i}}={{\dot \omega }_i} + {m_{P_i}}{{\dot P}_i}=u_{f_i},
\label{eq3}\\
& {{\dot V}_{n_i}}={{\dot v}_{odi}} + {n_{Q_i}}{{\dot Q}_i}=u_{v_i},
\label{eq4}
\end{align}
where $u_{f_i}$ and $u_{v_i}$ are auxiliary control inputs. To synchronize the terminal frequency of each inverter to the reference value and contain the terminal voltage of each inverter within the accepted range, the leader-follower containment-based secondary control is adopted \cite{yz052802}. The local cooperative frequency and voltage control protocols, using the relative information with respect to the neighboring inverters and the leaders, are given by
\begin{equation}
\begin{gathered}
  {u_{{f_i}}} = {c_{f_i}}\left( {\sum\limits_{j \in \mathscr{F}} {a_{ij}\left( {{\omega _j} - {\omega _i}} \right)} } \right. + \sum\limits_{k \in \mathscr{L}} {g_{ik}\left( {{\omega _k} - {\omega _i}} \right)}  \hfill \\
  \left. {\quad \quad  + \sum\limits_{j \in \mathscr{F}} {a_{ij}\left( {{m_{{P_j}}}{P_j} - {m_{{P_i}}}{P_i}} \right)} } \right), \hfill \\ 
\end{gathered}
\label{eq5}
\end{equation}
\begin{equation}
\begin{gathered}
  {u_{{v_i}}} = {c_{v_i}}\left( {\sum\limits_{j \in \mathscr{F}} {a_{ij}\left( {v _{odj} - v _{odi}} \right)} } \right. + \sum\limits_{k \in \mathscr{L}} {g_{ik}\left( {{v _k} - v _{odi}} \right)}  \hfill \\
  \left. {\quad \quad  + \sum\limits_{j \in \mathscr{F}} {a_{ij}\left( {{n_{{Q_j}}}{Q_j} - {n_{{Q_i}}}{Q_i}} \right)} } \right), \hfill \\ 
\end{gathered}
\label{eq6}
\end{equation}
% where ${c_{f_i} }, {c_{v_i} }$ are positive constant coupling gains. ${\omega _k}$ and $v_k$ are the frequency and voltage reference values of the $k^{th}$ leader, respectively. The frequency reference for both leaders is set as ${\omega _{\operatorname{ref}}}$. Given that the two frequency references coincide, the containment problem simplifies to a regulation problem in this specific case. The upper and lower leaders have their voltage reference values set as ${v_{\operatorname{ref}}^u}$ and ${v_{\operatorname{ref}}^l}$, respectively. {\color{blue}\sout{Specifically, the voltage objective is to attain voltage convergence of the local inverter to a small neighborhood around the range spanned by the upper and lower voltage reference values. Furthermore, the frequency objective is to achieve frequency convergence of the local inverter to a small neighborhood around the frequency reference value.}}
% {\color{RawSienna}\sout{Specifically, the frequency of local inverter converges to a small neighborhood around the frequency reference value.}}
% {\color{RawSienna}\sout{Specifically, the voltage of local inverter converges to a small neighborhood around the range spanned by the upper and lower voltage reference values.}}

% (The containment objectives of frequency and voltage). 
where $c_{f_i}$ and $c_{v_i}$ are constant gains. The setpoints for the primary-level droop control, $\omega_{n_i}$ and ${V_{n_i}}$, are, then, computed from $u_{f_i}$ and $u_{v_i}$ as 
\begin{align}
&{\omega _{n_i}} = \int {u_{f_i}} \operatorname{d} t,\label{eq7}\\
&{V _{n_i}} = \int {u_{v_i}} \operatorname{d} t.
\label{eq8}
\end{align}

Using \eqref{eq5} and \eqref{eq6} to rewrite \eqref{eq3} and \eqref{eq4} yields
\begin{align}
{{\dot \omega }_{{n_i}}} &= {c_{f_i}}\left( {\sum\limits_{j \in \mathscr{F}} {a_{ij}\left( {{\omega _{{n_j}}} - {\omega _{{n_i}}}} \right)}  + \sum\limits_{k \in \mathscr{L}} {g_{ik}\left( {{\omega _{n_k}} - {\omega _{{n_i}}}} \right)} } \right),\label{eq9}
\end{align}
\begin{equation}
{{\dot V}_{{n_i}}} = {c_{v_i}}\left( {\sum\limits_{j \in \mathscr{F}} {a_{ij}\left( {{V_{{n_j}}} - {V_{{n_i}}}} \right)}  + \sum\limits_{k \in \mathscr{L}} {g_{ik}\left( {V_{n_k} - {V_{{n_i}}}} \right)} } \right),
\label{eq11}
\end{equation}
where ${\omega _{{n_k}}}={\omega _k} + {m_{{P_i}}}{P_i}$ and $V_{n_k}={v_k + {n_{{Q_i}}}{Q_i}}$. Define ${\Phi _k} = \frac{1}{2}{\mathcal{L}} + \mathcal{G}_k$. Then, the global forms of \eqref{eq9} and \eqref{eq11} are
\begin{equation}
{{\dot \omega }_n} =  - \operatorname{diag} \left( {{c_{{f_i}}}} \right)\sum\limits_{k \in \mathscr{L}} {{\Phi _k}\left( {{\omega _n} - {{\mathbf{1}}_N} \otimes {\omega _{{n_k}}}} \right),}
\label{eq12}
\end{equation}
\begin{equation}
\dot V_n =  - {\operatorname{diag} \left( {{c_{v_i}}} \right)}\sum\limits_{k \in \mathscr{L}} {\Phi _k\left( {V_n - {{\mathbf{1}}_N} \otimes {V_{n_k}}} \right),} 
\label{eq13}
\end{equation}
where $\omega_n= {[ {\omega_{n_1}^T,...,\omega_{n_N}^T} ]^T}$ and $V_n= {[ {V_{n_1}^T,...,V_{n_N}^T} ]^T}$. Define the global frequency and voltage containment error vectors as
\begin{equation}
{e_f} = {\omega _n} - {\left( {\sum\limits_{k \in \mathscr{L}} {{\Phi _k}} } \right)^{ - 1}}\sum\limits_{k \in \mathscr{L}} {{\Phi _k}\left( {{{\mathbf{1}}_N} \otimes {\omega _{{n_k}}}} \right)} ,
\label{eq14}
\end{equation}
\begin{equation}
{e_v} = V_n - {\left( {\sum\limits_{k \in \mathscr{L}} {{\Phi _k}} } \right)^{ - 1}}\sum\limits_{k \in \mathscr{L}} {{\Phi _k}\left( {{{\mathbf{1}}_N} \otimes V_{n_k}} \right)}.
\label{eq15}
\end{equation}

% \textbf{\textit{Definition} 1 \textit{(Secondary frequency containment control objective)}:} The secondary frequency control objective is to make the local frequency of each inverter converge to the range of the two frequency references issued by the upper and lower leaders. Since these two reference values are identical, the frequency regulation is achieved.

% \textbf{\textit{Definition} 2 \textit{(Secondary voltage containment control objective)}:} The secondary voltage containment control objective is to make each inverter voltage converge to the range spanned by the two references of the upper and lower leaders.

The following assumption is needed for the communication graph topology to guarantee cooperative consensus.

\textbf{\textit{Assumption} 1:}
%The communication graph $\mathscr{G}$ includes a directed path from, at least, one leader to each inverter.
There exists a directed path from each leader to each inverter.

\textbf{\textit{Lemma} 1 (\cite{yz052802}):} Suppose Assumption 1 holds, $\sum\nolimits_{k \in \mathscr{L}} {\Phi _k} $ is non-singular and positive-definite. The frequency and voltage containment control objectives are achieved if $\mathop {\lim }\limits_{t \to \infty } e_f \left( t \right) = 0$ and $\mathop {\lim }\limits_{t \to \infty } e_v \left( t \right) = 0$, respectively.

\section{Problem Formulation}
In this section, we formulate the attack-resilient defense problems for the secondary frequency and voltage control of an AC microgrid. Specifically, we introduce generally unbounded FDI attacks on the local input channels of the frequency and voltage control loops, which results in the following modifications to \eqref{eq9} and \eqref{eq11} 
\begin{equation}
\begin{gathered}
  {{\dot \omega }_{n_i}} = {c_{{f_i}}}\left( {\sum\limits_{j \in \mathscr{F}} {{a_{ij}}\left( {{\omega _{{n_j}}} - {\omega _{{n_i}}}} \right)}  + \sum\limits_{k \in \mathscr{L}} {{g_{ik}}\left( {{\omega _{{n_k}}} - {\omega _{{n_i}}}} \right)} } \right) \hfill \\
  \quad \quad + \Delta_{f_i}, \hfill \\ 
\end{gathered}
\label{eq16}
\end{equation}
\begin{equation}
\begin{gathered}
  {{\dot V}_{n_i}} ={c_{v_i}}\left( {\sum\limits_{j \in \mathscr{F}} {{a_{ij}}\left( {{V_{{n_j}}} - {V_{{n_i}}}} \right)}  + \sum\limits_{k \in \mathscr{L}} {{g_{ik}}\left( {{V_{{n_k}}} - {V_{{n_i}}}} \right)} } \right) \hfill \\
  \quad \quad  + \Delta_{v_i}, \hfill \\ 
\end{gathered}
\label{eq17}
\end{equation}
where $\Delta_{f_i}$ and $\Delta_{v_i}$ denote the unbounded attack signals injected to the input channels of frequency and voltage control loops at the $i^{th}$ inverter, respectively.

%We use $\varpi _{{f_i}}^{\left( q \right)}$ and $\varpi _{{v_i}}^{\left( q \right)}$ to denote the $q$th derivative of ${\varpi _{{f_i}}}$ and ${\varpi _{{v_i}}}$, respectively.

\textbf{\textit{Assumption} 2:}
$\Delta_{f_i}(t)\in C^\gamma$ and $\Delta_{v_i}(t)\in C^\gamma$. $\left|\frac{\operatorname{d}^{\gamma}}{\operatorname{d}t^{\gamma}}\Delta_{f_i}\right| \leqslant \kappa_{f_i}$ and $\left|\frac{\operatorname{d}^{\gamma}}{\operatorname{d}t^{\gamma}}\Delta_{v_i}\right| \leqslant \kappa_{f_i}$, where $\kappa_{f_i}$ and $\kappa_{v_i}$ are positive constants.

%\textbf{\textit{Remark} 1:} 
%Assumption 2 is reasonable since attack signals, with an excessively-large change in values, could be easily detected in practice.

Since $\Delta_{f_i}$ and $\Delta_{v_i}$ are unbounded, conventional cooperative control protocols fail to regulate the frequency and contain the voltages within the accepted range. One then needs attack-resilient defense strategies to preserve the frequency regulation and voltage containment performances and assure the closed-loop stability. The following convergence definition is needed. 

\textbf{\textit{Definition} 1 (\cite{yz053105}):} 
Signal $x(t)$ is UUB with an ultimate bound $b$, if there exist positive constants $b$ and $c$, independent of ${t_0} \geq 0$, and for every $a \in \left( {0,c} \right)$, there exist $t_1 = t_1 \left( {a,b} \right) \geq 0$, independent of $t_0$, such that $\left\| {x\left( {{t_0}} \right)} \right\| \leq a\Rightarrow \left\| {x\left( t \right)} \right\| \leq b, \forall t \geq {t_0} + {t_1}$.

Now, we introduce the following attack-resilient defense problems for the secondary frequency and voltage control loops.

\textbf{\textit{Definition} 2 \textit{(Attack-resilient Frequency Defense Problem)}:}
The aim is to design an input control signal $u_{f_i}$, as delineated in equation \eqref{eq3}, for each inverter, such that the global frequency containment error $e_f$, as specified in equation \eqref{eq14}, remains UUB in the face of generally unbounded attacks on the local frequency control loop.

\textbf{\textit{Definition} 3 \textit{(Attack-resilient Voltage Defense Problem)}:}
The aim is to design an input control signal $u_{v_i}$, as delineated in equation \eqref{eq4}, for each inverter, such that the global voltage containment error $e_v$, as defined in equation \eqref{eq15}, remains UUB in the face of generally unbounded attacks on the local voltage control loop. 

\section{Fully Distributed Attack-resilient Defense strategies Design}

\begin{figure}[!t]
\centering
\includegraphics[width=3.5in]{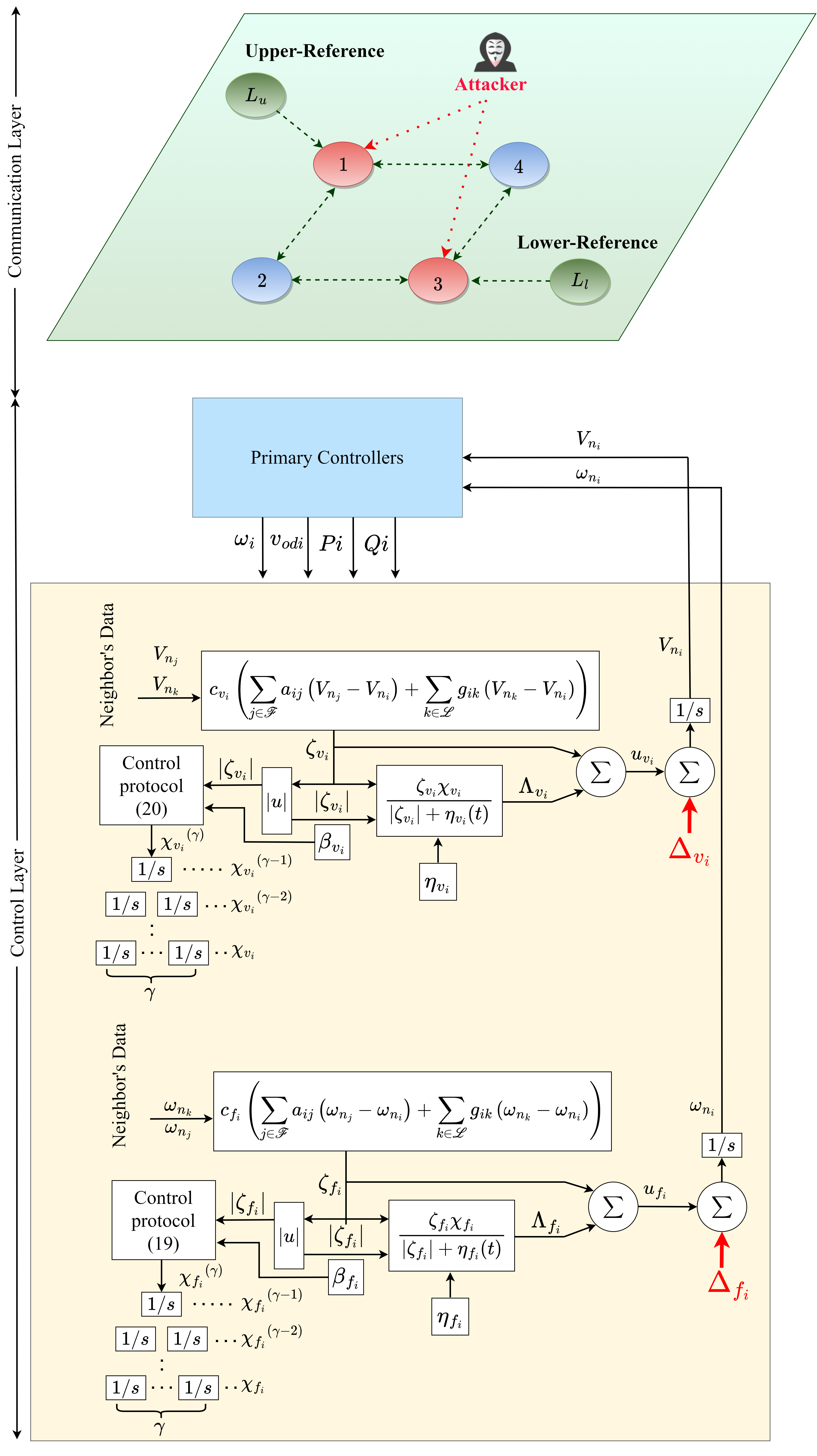}
\caption{Communication layer among inverters, and the proposed attack-resilient secondary defense strategies for an inverter.}
\label{FIG2}
\end{figure}

We propose the following fully distributed attack-resilient defense strategies to solve the attack-resilient frequency and voltage defense problems. For convenience, denote
\begin{equation}
{\zeta _{{f_i}}} = c_{f_i}\left(\sum\limits_{j \in \mathscr{F}} {{a_{ij}}\left( {{\omega _{{n_j}}} - {\omega _{{n_i}}}} \right)}  + \sum\limits_{k \in \mathscr{L}} {{g_{ik}}\left( {{\omega _{{n_k}}} - {\omega _{{n_i}}}} \right)}\right) ,
\label{eq18}
\end{equation}
\begin{equation}
{\zeta _{v_i}} = c_{v_i}\left(\sum\limits_{j \in \mathscr{F}} {{a_{ij}}\left( {{V_{{n_j}}} - {V_{{n_i}}}} \right)}  + \sum\limits_{k \in \mathscr{L}} {{g_{ik}}\left( {{V_{{n_k}}} - {V_{{n_i}}}} \right)}\right).
\label{eq19}
\end{equation}

Then, we present the following attack-resilient defense strategies for both frequency and voltage control loops
\begin{equation}
\left\{ \begin{gathered}
  u_{f_i} = {\zeta _{f_i} + {\Lambda}_{f_i}}, \hfill \\
  {{\Lambda}_{f_i}} = \frac{\zeta_{f_i}\chi_{f_i}}{\lvert\zeta_{f_i}\rvert + \eta_{f_i}}, \hfill \\ 
  {\chi^{(\gamma)}_{f_i}} = \beta_{f_i}\lvert\zeta_{f_i}\rvert.
\end{gathered}  \right.
\label{eq20}
\end{equation}
\begin{equation}
\left\{ \begin{gathered}
  u_{v_i} = {\zeta _{v_i} + {\Lambda}_{v_i}}, \hfill \\
  {{\Lambda}_{v_i}} = \frac{\zeta_{v_i}\chi_{v_i}}{\lvert\zeta_{v_i}\rvert + \eta_{v_i}}, \hfill \\ 
  {\chi^{(\gamma)}_{v_i}} = \beta_{v_i}\lvert\zeta_{v_i}\rvert.
\end{gathered}  \right.
\label{eq21}
\end{equation}
where $\eta_{f_i}$ and $\eta_{v_i}$ are positive exponentially decaying functions, ${\Lambda}_{f_i}$ and ${\Lambda}_{v_i}$ are compensational signals, $\chi_{f_i}$ and $\chi_{v_i}$ are adaptively tuned parameters, the adaptation gains $\beta_{f_i}$ and $\beta_{v_i}$ are given positive constants. The initial values of both $\chi_{f_i}$ and $\chi_{v_i}$ are positive. Figure~\ref{FIG2} shows the communication network among inverters and the proposed secondary defense strategies for an inverter.

\textbf{\textit{Theorem} 1:} Under Assumptions 1 and 2, and given the implementation of the cooperative attack-resilient voltage defense strategies as delineated in equations \eqref{eq18} and \eqref{eq20}, the error $e_f$, defined in equation \eqref{eq14}, is UUB, i.e., the attack-resilient frequency defense problem is solved. Additionally, it is observed that by properly adjusting the value of ${\beta _{f_i}}$ as prescribed in equation \eqref{eq20}, the ultimate bound of $e_f$ is reduced to an arbitrarily small value.

\textbf{\textit{Proof}:} 
Combining \eqref{eq9}, \eqref{eq12}, \eqref{eq16}, \eqref{eq18} and \eqref{eq20} yields the global form:

\begin{align}
\dot{\zeta}_f& = -\left(\sum_{k \in \mathscr{L}} \Phi_k\right)\operatorname{diag}\left(c_{f_i}\right)\dot \omega_n \nonumber\\
& = - \left(\sum_{k \in \mathscr{L}} \Phi_k\right)\operatorname{diag}\left(c_{f_i}\right)\big(\zeta_f +\Delta_f +\Lambda_f\big),
\label{eq22}
\end{align}

where $\zeta_f= [ \zeta_{f_i}^T,...,\zeta_{f_N}^T ]^T, \Delta_f= [ \Delta_{f_i}^T,...,\Delta_{f_N}^T ]^T$ and $\Lambda_f= [ \Lambda_{f_i}^T,...,\Lambda_{f_N}^T ]^T$.

Consider the following Lyapunov function candidate
\begin{align}
E & =\frac{1}{2} \zeta_f^{T}\left(\sum_{k \in \mathscr{L}} \Phi_k\right)^{-1} \zeta_f.
\label{eq23}
\end{align}
Its time derivative is
\begin{align}
\dot{E} & =\frac{1}{2} \times 2 \zeta_f^{T}\left(\sum_{k \in \mathscr{L}} \Phi_k\right)^{-1} \dot{\zeta}_f \nonumber\\
& =-\zeta_f^{T}\left(\sum_{k \in \mathscr{L}} \Phi_k\right)^{-1}\left(\sum_{k \in \mathscr{L}} \Phi_k\right) \operatorname{diag}\left(c_{f_i}\right)\big(\zeta_f\nonumber\\
&+\Delta_f +\Lambda_f\big) \nonumber\\
& \leqslant-\sigma_{\min }\left(\operatorname{diag}\left(c_{f_i}\right)\right)\left\|\zeta_f\right\|^2-\operatorname{diag}\left(c_{f_i}\right) \zeta_f^{T} \Delta_f\nonumber\\
&-\operatorname{diag}\left(c_{f_i}\right) \zeta_f^{T} \Lambda_f
\nonumber\\
& = -\sigma_{\min }\left(\operatorname{diag}\left(c_{f_i}\right)\right)\left\|\zeta_f\right\|^2-\operatorname{diag}\left(c_{f_i}\right)\sum\limits_{i \in \mathscr{F}} \big(\zeta_{f_i} \Delta_{f_i}\big)\nonumber\\
&-\operatorname{diag}\left(c_{f_i}\right)\sum\limits_{i \in \mathscr{F}} \big(\zeta_{f_i} \Lambda_{f_i}\big)
\nonumber\\
& \leqslant -\sigma_{\min }\left(\operatorname{diag}\left(c_{f_i}\right)\right)\left\|\zeta_f\right\|^2 + \operatorname{diag}\left(c_{f_i}\right)\sum\limits_{i \in \mathscr{F}} \big|\zeta_{f_i} \big|\big|\Delta_{f_i}\big|\nonumber\\
&-\operatorname{diag}\left(c_{f_i}\right)\sum\limits_{i \in \mathscr{F}} \big(\zeta_{f_i} \Lambda_{f_i}\big).
\label{eq24}
\end{align}
Upon substituting $\Lambda_{f_i}$ from Equation \eqref{eq20}, the final two terms on the right-hand side of Equation \eqref{eq24} are transformed as follows
\begin{align}
& \operatorname{diag}\left(c_{f_i}\right)\sum\limits_{i \in \mathscr{F}} \big|\zeta_{f_i} \big|\big|\Delta_{f_i}\big|-\operatorname{diag}\left(c_{f_i}\right)\sum\limits_{i \in \mathscr{F}} \big(\zeta_{f_i} \Lambda_{f_i}\big) \nonumber\\
& = \operatorname{diag}\left(c_{f_i}\right)\sum\limits_{i \in \mathscr{F}}\left( \big|\zeta_{f_i} \big|\big|\Delta_{f_i}\big|- \frac{\big|\zeta_{f_i}\big|^2 \chi_{f_i}}{\left|\zeta_{f_i}\right|+\eta_{f_i}}\right)
\nonumber\\
& =  \operatorname{diag}\left(c_{f_i}\right)\sum\limits_{i \in \mathscr{F}}\left( \frac{\big|\zeta_{f_i}\big|^2\big|\Delta_{f_i}\big| + \big|\zeta_{f_i} \big|\big|\Delta_{f_i}\big|\eta_{f_i} - \big|\zeta_{f_i}\big|^2 \chi_{f_i} }{\left|\zeta_{f_i}\right|+\eta_{f_i}}\right)
\nonumber\\
& =  \operatorname{diag}\left(c_{f_i}\right)\sum\limits_{i \in \mathscr{F}}\left(\big|\zeta_{f_i}\big| \frac{\big|\zeta_{f_i}\big|\big(\big|\Delta_{f_i}\big| - \chi_{f_i}\big) + \big|\Delta_{f_i}\big|\eta_{f_i}}{\left|\zeta_{f_i}\right|+\eta_{f_i}}\right).
\label{eq25}
\end{align}

Since $\eta_{f_i}$ is an exponentially decaying function, based on Assumption 2, $\lim_{{t \to \infty}}\big|\Delta_{f_i}\big|\eta_{f_i} = 0$. From \eqref{eq20}, when $\left|\zeta_{f_i}\right| \geqslant \left|\frac{\operatorname{d}^{\gamma}}{\operatorname{d}t^{\gamma}}\Delta_{f_i}\right|/ \beta_{f_i}$, i.e., $ \left|\zeta_{f_i}\right| \geqslant\kappa_{f_i} / \beta_{f_i}$, $\exists t_1$, such that $\chi_{f_i} \geqslant \left|\Delta_{f_i}\right|$.
This suggests that $\exists t_2 > t_1$ such that 
\begin{align}
&\operatorname{diag}\left(c_{f_i}\right)\sum\limits_{i \in \mathscr{F}} \big|\zeta_{f_i} \big|\big|\Delta_{f_i}\big|-\operatorname{diag}\left(c_{f_i}\right)\sum\limits_{i \in \mathscr{F}} \big(\zeta_{f_i} \Lambda_{f_i}\big) \leqslant 0,\nonumber\\ &\forall t \geqslant t_2.
\label{eq26}
\end{align}

Considering \eqref{eq24}, \eqref{eq25} and \eqref{eq26} yields 
\begin{align}
\dot E \leqslant 0,\; 
\forall \left|\zeta_{f_i}\right| \geqslant\kappa_{f_i} / \beta_{f_i} ,\forall t\geqslant t_2.
\label{eq27}
\end{align}
Hence, $\zeta_f$ is UUB. From Theorem 4.18 of \cite{yz053105}, while the system stability is maintained, the larger the value of the adaptation gain $\beta_{f_i}$, the smaller the ultimate bound. Note that $\zeta_f = \sum\limits_{k \in \mathscr{L}}\Phi_k e_f$. Hence $e_f$ is also bounded.

\textbf{\textit{Theorem} 2:} Under Assumptions 1 and 2, the cooperative attack-resilient voltage defense strategies described by equations \eqref{eq19} and \eqref{eq21} ensure that $e_v$ in equation \eqref{eq15} is UUB, i.e., the attack-resilient voltage defense problem is solved. Moreover, by properly adjusting the adaptation gain ${\beta_{v_i}}$ in equation \eqref{eq21} the ultimate bound of $e_v$ is set arbitrarily small.

\textbf{\textit{Proof}:} 
The approach used to prove Theorem 2 mirrors that of Theorem 1.
\hfill\(\blacksquare\)

% \textbf{\textit{Remark} 2:} 
% Compared to \cite{yz052802}, the proposed control protocols \eqref{eq18}-\eqref{eq21} have the following merits: (i) Local observers with additional communication information flow were constructed in \cite{yz052802} to estimate the actual state measurements. This, however, could introduce additional computational complexity. Moreover, the additional communication channels for exchanging observer states could potentially increase the system vulnerability to malicious cyber attacks; (ii) While both \cite{yz052802} and this letter preserve the UUB convergences for both frequency and voltage terms, in this letter, the ultimate bound can be reduced by properly increasing the adaptive tuning parameters. 

\section{Validation and Results}%Simulation Results and Discussion} 

\begin{figure}[!t]
\centering
\includegraphics[width=3.2in]{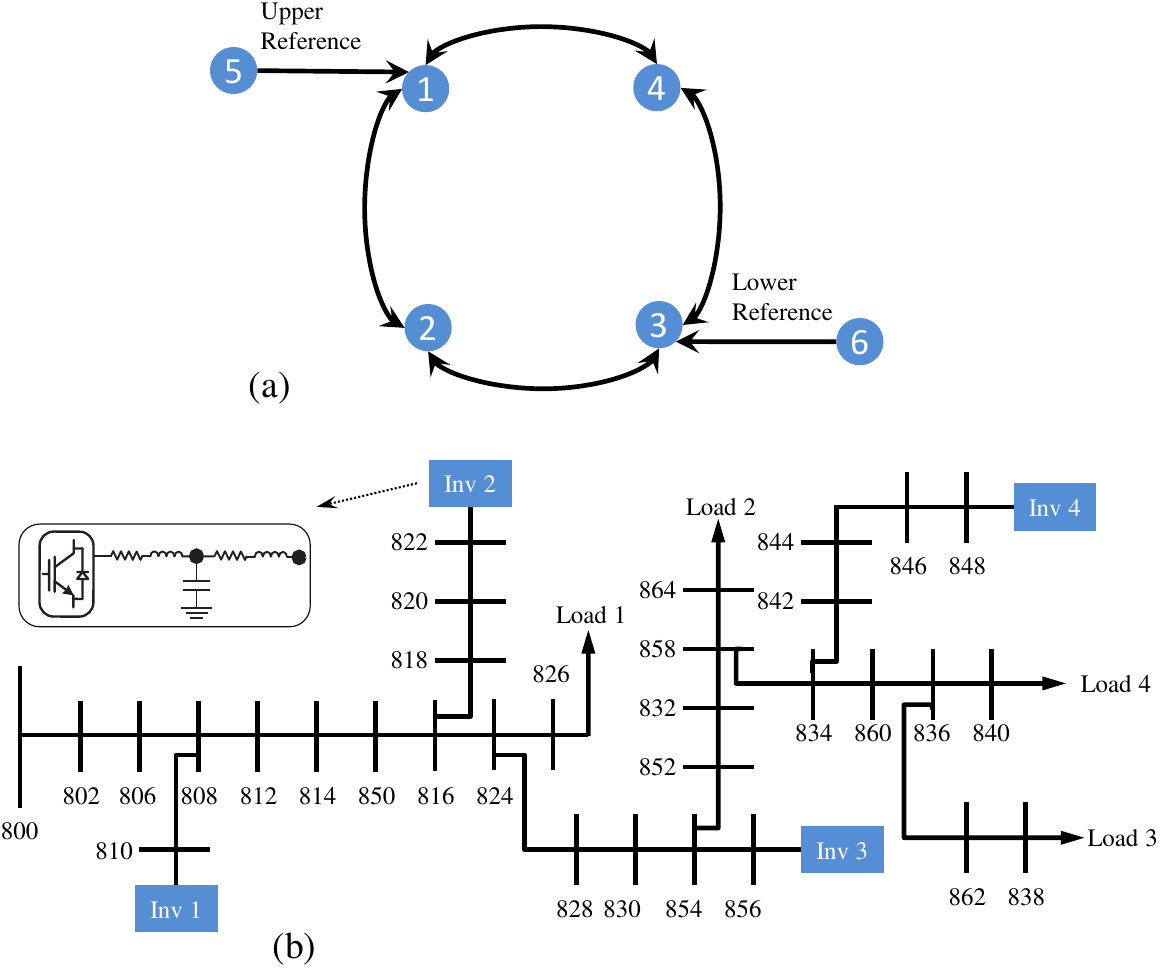}
\caption{Cyber-physical microgrid system: (a) Communication graph topology among four inverters and two leaders (references), (b) IEEE 34-bus system with four inverters.}
\label{FIG3}
\end{figure}

The proposed fully distributed attack-resilient secondary defense strategies are validated on an IEEE 34-bus feeder system, islanded at bus 800 and incorporating four inverter-based DERs and two leaders (references), as shown in Fig. \ref{FIG3}. This section presents three different case studies to show the effectiveness of the proposed theoretical results. 

Specifications of inverters and their grid-interconnections are adopted from \cite{yz052801} and \cite{28}, respectively. All inverters have the same power ratings. The inverter droop gains are set as $m_{P_1}=m_{P_2}=9.4\times {10^{ - 5}}$, $m_{P_3}=m_{P_4}=18.8\times {10^{ - 5}}$, $n_{Q_1}=n_{Q_2}=1.3\times {10^{ - 3}}$, and $n_{Q_3}=n_{Q_4}=2.6\times {10^{ - 3}}$. The inverters communicate on a bidirectional communication network with the adjacency matrix of $\mathcal{A}=[0~1~0~1;1~0~1~0;0~1~0~1;1~0~1~0]$. The pinning gains are $g_{15}=g_{36}=1$. The frequency reference, upper voltage reference, and lower voltage reference are $60 \,\operatorname{Hz}$,  $350\,\operatorname{V}$, and  $330\,\operatorname{V}$, respectively.

\subsection{Case Study I: Resilience against Generally Unbounded Attacks}

In this case study, the unbounded attack injections to the frequency and voltage control loops are set as $\Delta_{f_1}=0.5t^2, \Delta_{f_2}=0.4t^2, \Delta_{f_3}=0.5t^2, \Delta_{f_4}=0.3t^2$ and $\Delta_{v_1}=0.5t^2, \Delta_{v_2}=t^2, \Delta_{v_3}=0.3t^2, \Delta_{v_4}=0.4t^2$, respectively.
%$\Delta_{f_i}=(0.5t^2,0.4t^2,0.5t^2,0.3t^2), i = 1,2,3,4$ and $\Delta_{v_i}=(0.5t^2,t^2,0.3t^2,0.4t^2), i = 1,2,3,4$, respectively. 
The performance of the attack-resilient defense strategies, \eqref{eq18}-\eqref{eq21}, is compared with the conventional secondary control method in \eqref{eq5} and \eqref{eq6}. The constant gains for the conventional control protocols are set as $c_{f_i}=20, c_{v_i}=10, i=1,2,3,4$. The adaptation gains for the resilient defense strategies are set as ${\beta_{v_i}}=20,  {\beta_{f_i}}=350, i=1,2,3,4.$ $\eta_{v_i}$ and $\eta_{f_i}$ are specifies as $e^{-\alpha_{v_i}}$ and $e^{-\alpha_{f_i}}$, where ${\alpha_{v_i}}={ \alpha_{f_i}}=0.01, i=1,2,3,4$.

\begin{figure}[!ht]
\centering
{\includegraphics[width=3.4in]{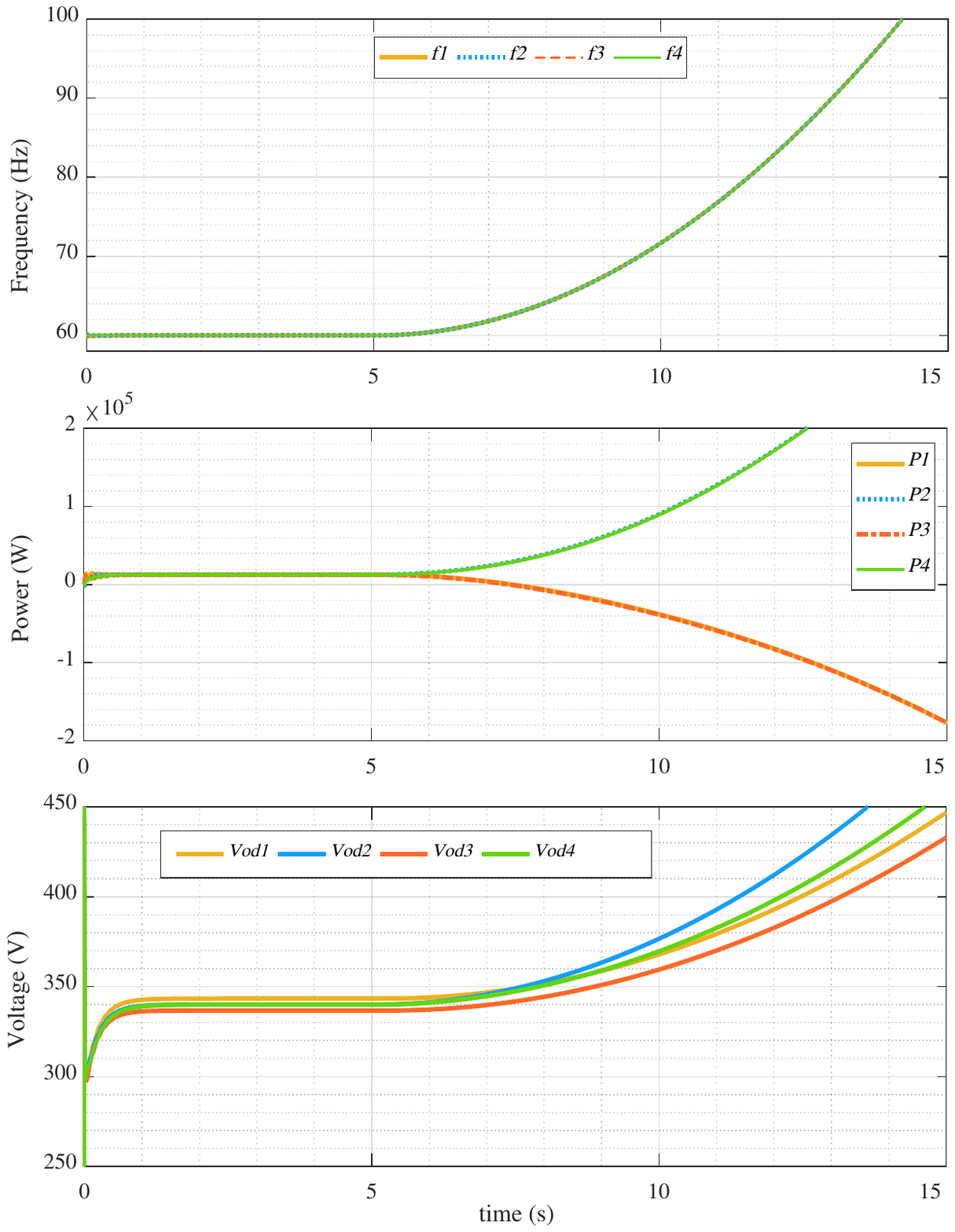}}
\caption{Performance of the conventional secondary control under unbounded attack signals: frequency performance, active power of inverters and voltage performance.}
\label{FIG40}
\end{figure}

\begin{figure}[!ht]
\centering
{\includegraphics[width=3.6in]{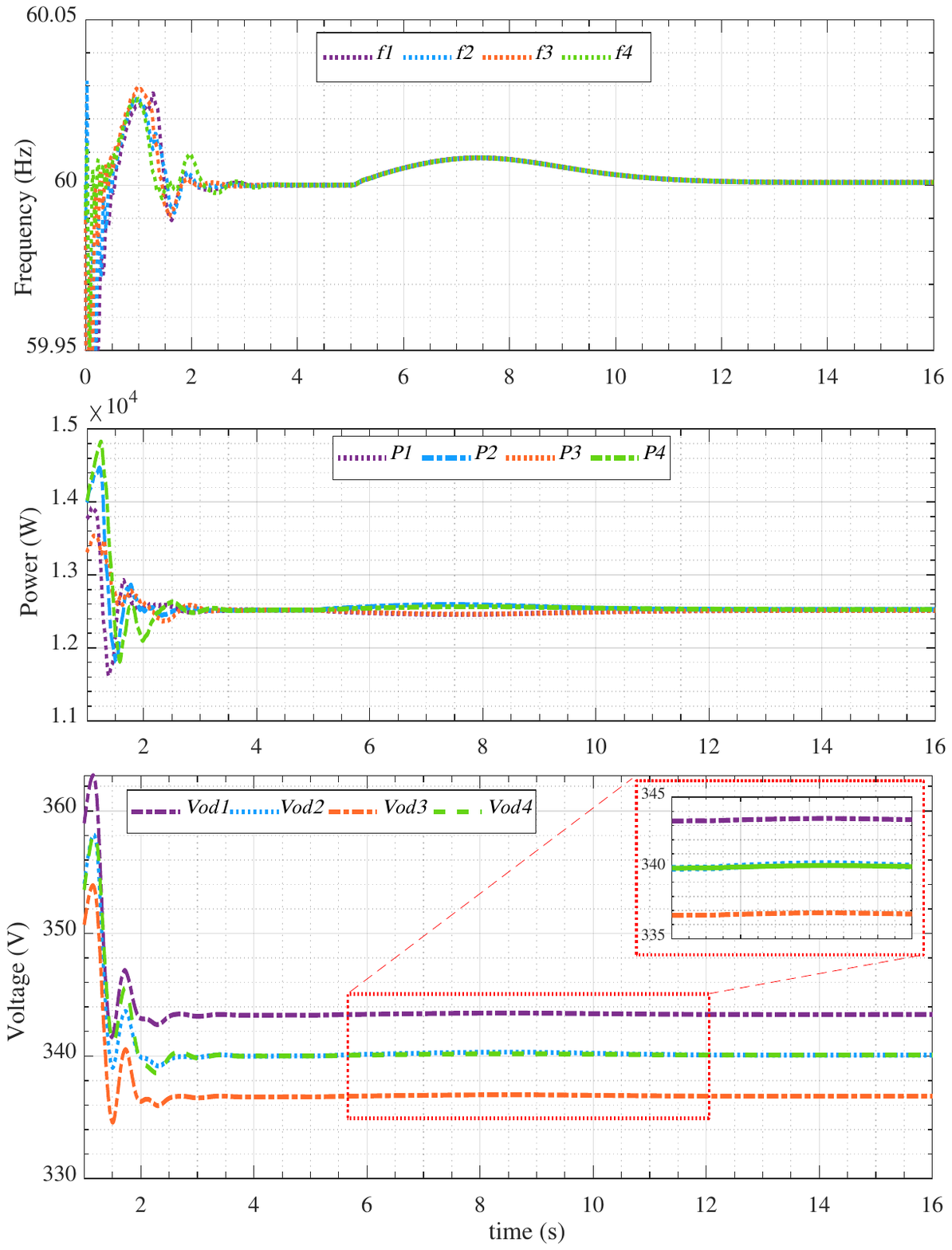}}
\caption{Performance of the proposed fully distributed attack-resilient secondary defense strategies under unbounded attack signals:  frequency performance, active power of inverters and voltage performance.}
\label{FIG50}
\end{figure}

Figures \ref{FIG40} and \ref{FIG50} compare the voltage and frequency responses against unbounded attacks using the conventional and the proposed resilient secondary defense strategies. As seen, after initiating the attack injections at $t=5 \,\mathrm{s}$, both voltage and frequency diverge. Therefore, the conventional method fails to preserve the system stability. Besides, proper active power sharing is not accomplished after initiating the attack injections at $t=5 \,\mathrm{s}$ using the conventional secondary control approach.

In contrast, by utilizing the proposed attack-resilient secondary defense strategies, the voltage of each inverter converges to a value within the range $330–350\, \mathrm{V}$, the frequency converges to the reference value, $60 \,\mathrm{Hz}$ and the active powers converge to the same values reflecting the equally shared power. The aforementioned results validate that the proposed resilient defense strategies accomplish the UUB convergence on frequency regulation, achieve voltage containment and active power sharing simultaneously for multi-inverter-based AC microgrids, in the face of unbounded attacks.

\begin{figure}[!ht]
\centering
{\includegraphics[width=3.5in]{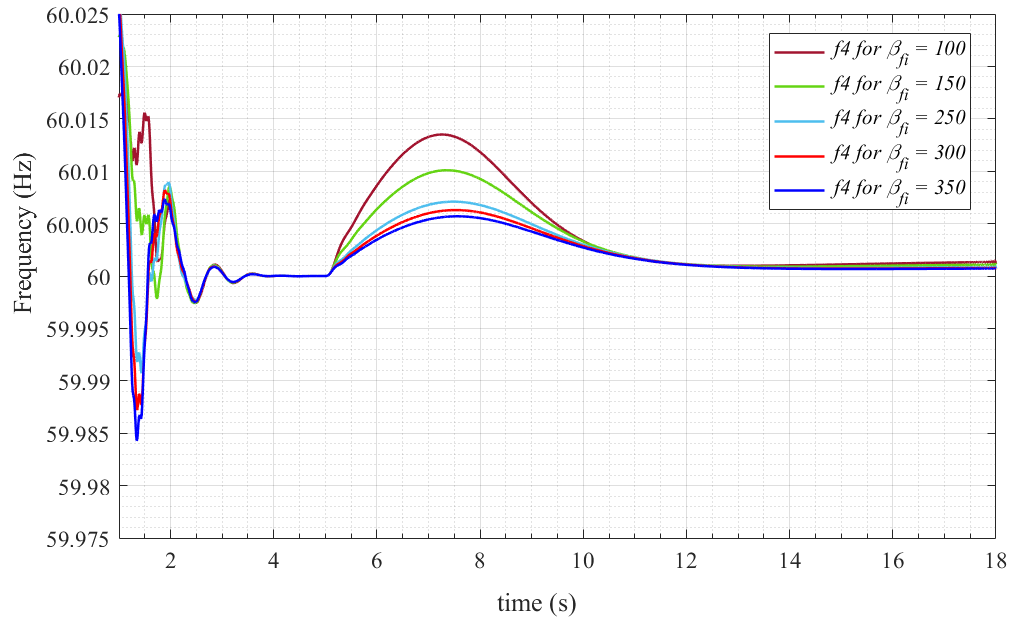}}
\caption{Comparative frequency performance under unbounded actuator attacks with different values of adaptation gain ${\beta _{f_i}}$.}
\label{FIG70}
\end{figure}

Furthermore, to validate that the ultimate bound of the UUB convergence is set to an arbitrarily small value by properly adjusting the adaptation gains, Figure \ref{FIG70} shows the frequency versus time for inverter 4, where the performances with different ${\beta _{f_i}}$ are illustrated with different colors. As seen, the ultimate bound is reduced and the transient response is improved by properly adjusting the value of ${\beta _{f_i}}$.

\subsection{Case Study II: Robustness against Load Step}
\begin{figure}[!ht]
\centering
{\includegraphics[width=3.6in]{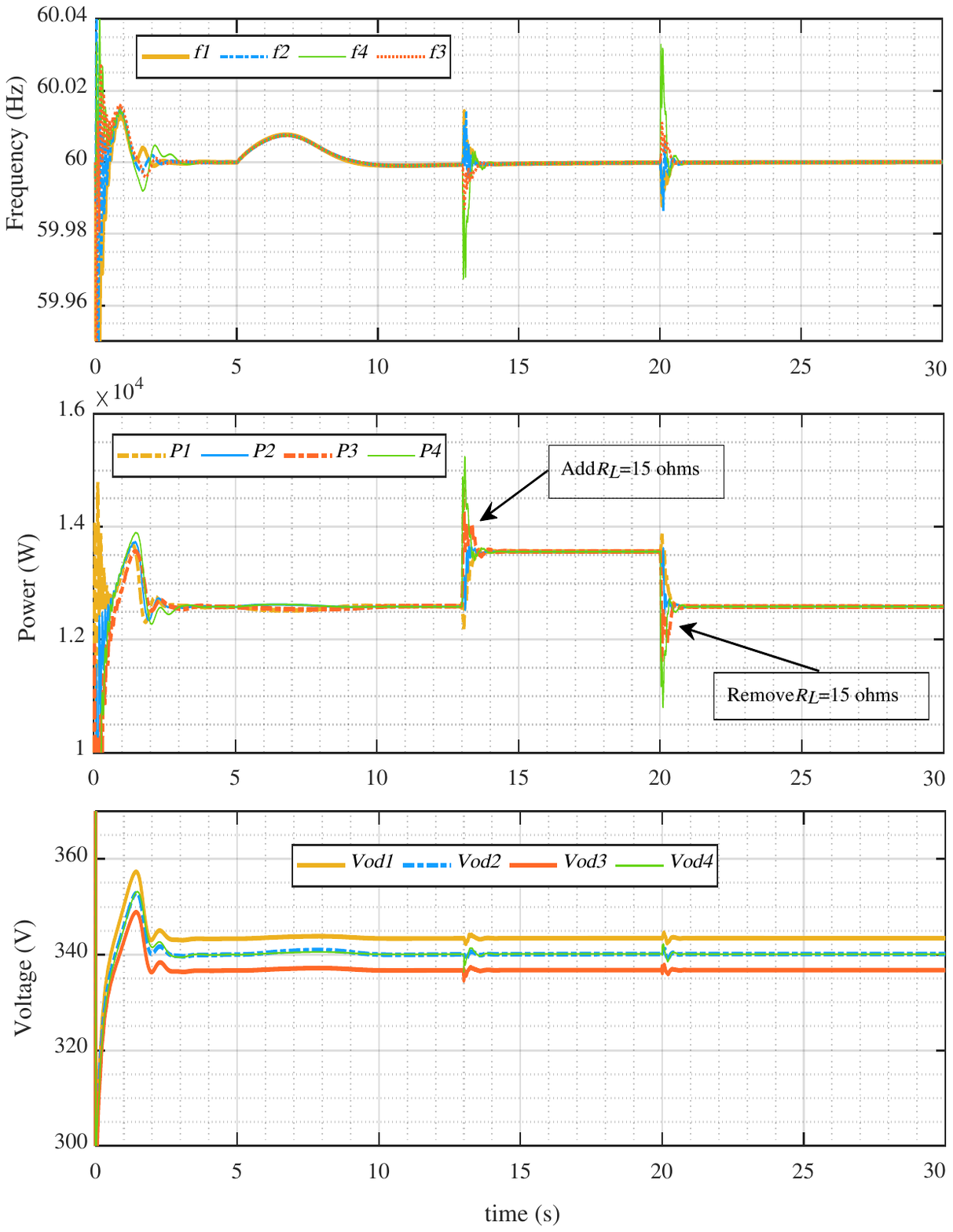}}
\caption{Performance of the proposed fully distributed attack-resilient secondary defense strategy in the case of unbounded attack signals and step load change: Voltage performance, Frequency performance, Active power performance.}
\label{FIG100}
\end{figure}

%distributed secondary control algorithm is verified by investigating the operation of the MG under adding an external load in step to the existing MG.
Following previous case study on resilience against generally unbounded attacks, the robustness of the proposed cyber-physical defense strategies against load changes in the face of the unbounded attacks is verified in this case study. The time span of the case study is from $0\,\mathrm{s}$ to $30\,\mathrm{s}$. Besides the attack injections initiated at $t=5\,\mathrm{s}$, additional load changes are initiated. Specifically, at $t = 13\,\mathrm{s}$, an additional load $R_L=15\,\Omega$ is introduced to the studied system, and this additional load is then eliminated at $t = 20\,\mathrm{s}$. Figure \ref{FIG100} shows the performance of the proposed fully distributed attack-resilient secondary defense strategy in the case of unbounded attack injections and step load change. As shown in Figure \ref{FIG100}, the proposed attack-resilient strategies maintains the UUB  regulation on frequency and accomplishes voltage containment of each inverter, as well as properly adjusts the active power shared in case of load change and unbounded attack injections. Therefore, the performance of the proposed resilient defense strategies does not deteriorate and is robust against load changes in the face of unbounded attacks.

\subsection{Case Study III: Robustness against Communication Link Failure}

\begin{figure}[!ht]
\centering
{\includegraphics[scale=0.2]{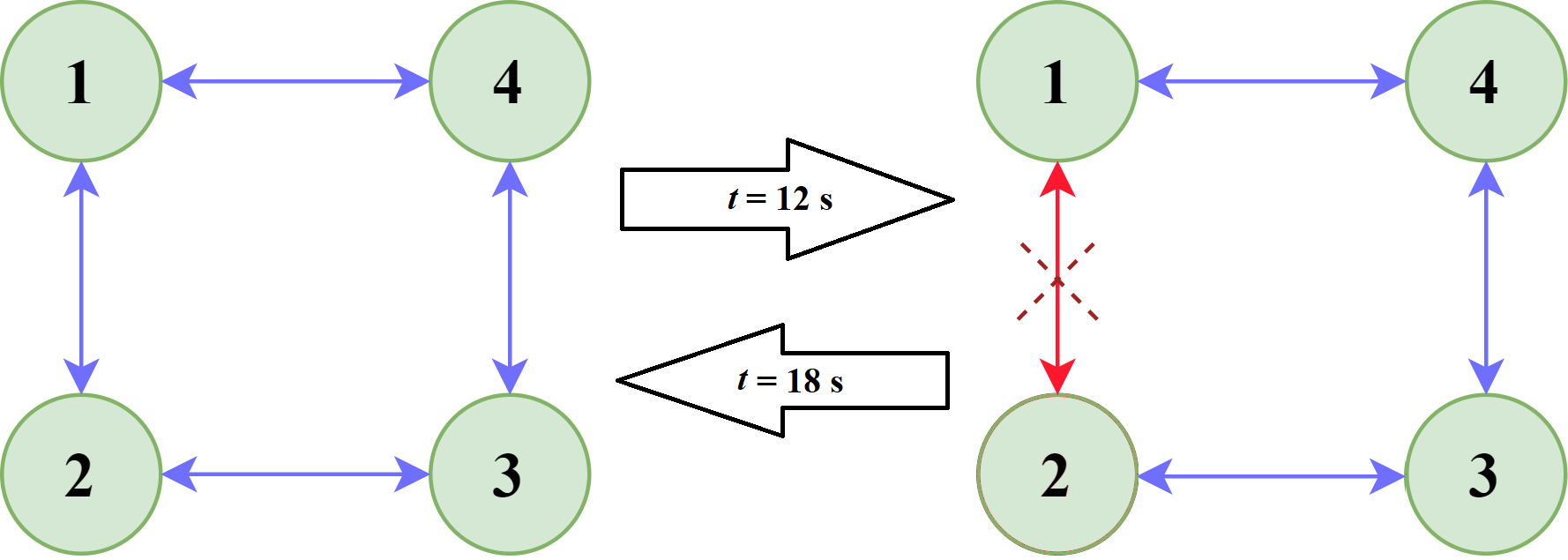}}
\caption{Communication topology of inverters in case study III.}
\label{FIG120}
\end{figure}

\begin{figure}[!ht]
\centering
{\includegraphics[width=3.6in]{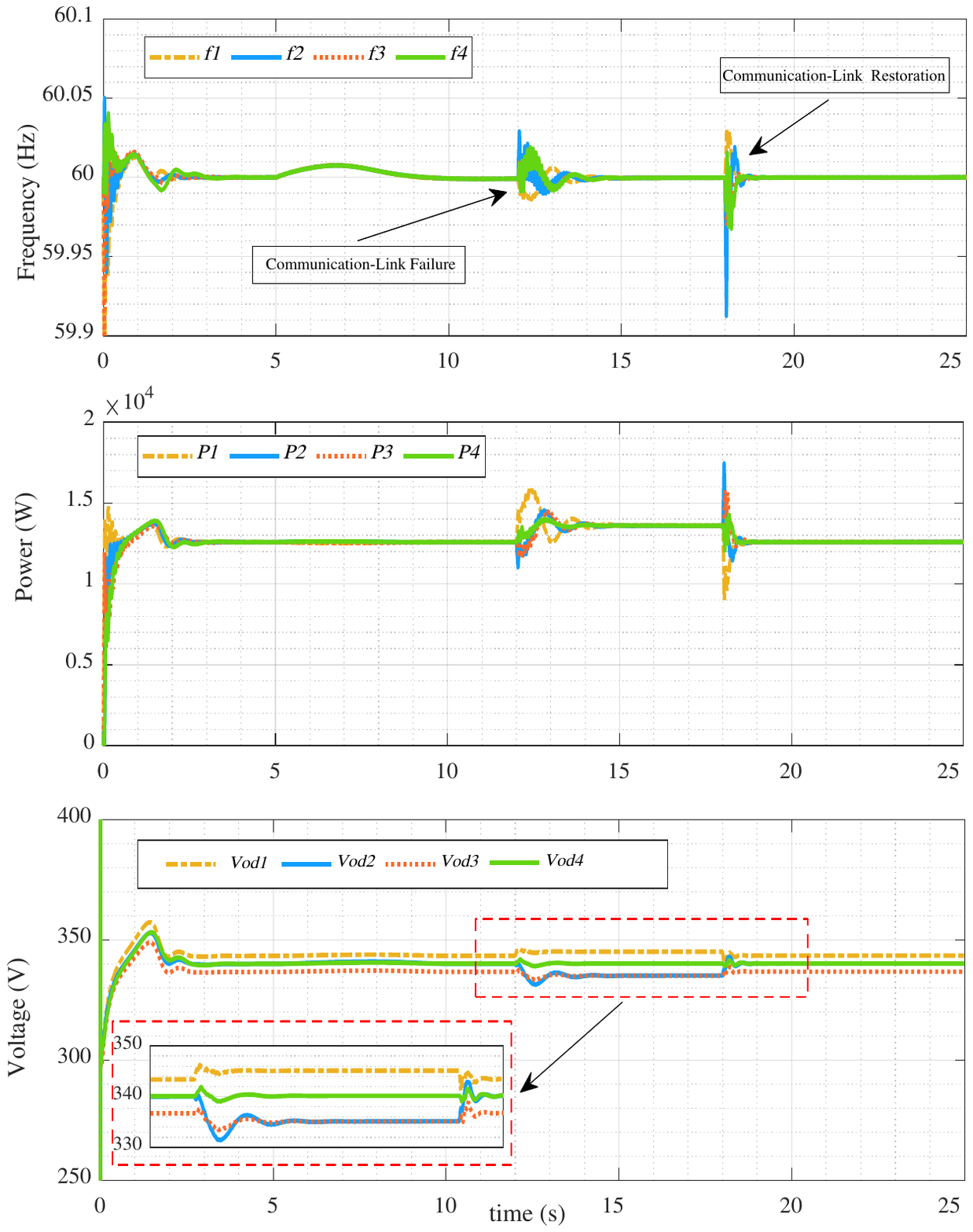}}
\caption{Performance of the proposed fully distributed attack-resilient secondary defense strategies in the case of unbounded attack signals and communication link failure: Voltage performance, Frequency performance, Active power performance.}
\label{FIG111}
\end{figure}

Finally, to demonstrate the robust performance of the proposed cyber-physical defense strategies under topological uncertainties of the microgrid caused by communication link failures, in addition to the initiation of the attack injections at $t=5\,\mathrm{s}$, the communication link between inverter 2 and inverter 1 is disabled at $t=12\,\mathrm{s}$ and restored at $t=18\,\mathrm{s}$ as illustrated in Fig. \ref{FIG120}. Figure \ref{FIG111} shows the performance of the proposed defense strategies. As seen, in the face of the communication link failures between inverter 2 and inverter 1 and the unbounded attacks, the proposed cyber-physical defense strategies maintains the UUB regulation on frequencies and accomplishes voltage containments, as well as properly adjusts the active powers shared.

%Therefore, validation results illustrate that our proposed cyber-physical defense strategies have plug-and-play capability.

%The communication topology changes according to Fig. 6, which ensures the communication network over the remained converters is strongly connected. If the i-th converter disconnects from the bus, then the link between the neighbors of the i-th converter is established as shown in Fig. 6. The trajectories of output currents and the DC bus voltage are shown in Fig. 7.

%\begin{figure}[!ht]
%\centering
%{\includegraphics[width=3.6in]{Plug.pdf}}
%\caption{Performance of the proposed fully distributed attack-resilient secondary defense strategy in the case of unbounded attack signals $\Delta_{f_i}$ and $\Delta_{v_i}$ and inverter $2$ disconnected: Voltage performance, Frequency performance, Active power performance.}
%\label{FIG110}
%\end{figure}

%\begin{figure}[!ht]
%\centering
%{\includegraphics[width=3.6in]{Inverter_Failure.pdf}}
%\caption{Performance of the proposed fully distributed attack-resilient secondary defense strategy in the case of unbounded attack signals $\Delta_{f_i}$ and $\Delta_{v_i}$ and inverter $2$ failure: Voltage performance, Frequency performance, Active power performance.}
%\label{FIG111}
%\end{figure}

%Figure \ref{FIG60} compares the frequency response for the proposed and the conventional methods. Under ideal conditions (no attacks), inverters frequencies synchronize to $f=60 \operatorname{Hz}$ using both control methods. Once the unbounded attack to frequency control loops is initiated at $t=4s$, the conventional method fails to preserve the system stability. By contrast, the proposed resilient method contains frequencies at a small neighborhood around 60 $\operatorname{Hz}$. Figure \ref{FIG5} shows that, without attacks, both methods share active powers among inverters based on their droop gains. After initiating the unbounded attacks to frequency control loops at $t=4s$, the active power performance from the conventional method becomes unstable. Meanwhile, the proposed method contains active powers in a small neighborhood around the value of properly shared powers. Figure \ref{FIG6} compares inverters voltages using both control methods. Without attacks, voltage values stay in the range of $330 \operatorname{V}$ to $350 \operatorname{V}$. After initiating the unbounded attacks to voltage control loops at $t=4s$, the voltages terms using the conventional method diverge, while those produced by the proposed method remain stable within $330 \operatorname{V} \sim 350 \operatorname{V}$. 

\section{Conclusion}
This paper has presented novel secondary cyber-physical defense strategies for multi-inverter AC microgrids against generally unbounded attacks on input channels of both frequency and voltage control loops. The proposed fully distributed cyber-physical defense strategies based on adaptive control techniques ensure the UUB stability of the closed-loop system by preserving the UUB consensus for frequency regulation and achieving voltage containment. Moreover, the ultimate bounds of convergence can be tuned by properly adjusting the adaptation gains, $\beta_{f_i}$ and $\beta_{v_i}$, in the adaptive tuning laws. The enhanced resilient performance of the proposed cyber-physical defense strategies has been verified using a modified IEEE 34-bus system.

\ifCLASSOPTIONcaptionsoff
  \newpage
\fi

\bibliographystyle{IEEEtran}

\bibliography{ref123}

\end{document}